\documentclass[aps,prc,preprint,groupedaddress,showpacs,floatfix]{revtex4}

\usepackage{graphicx}
\usepackage{amssymb}
\usepackage{bm}
\usepackage{color}

\newcommand{\be}{\begin{equation}}
\newcommand{\ee}{\end{equation}}
\newcommand{\bel}[1]{\be\label{#1}}
\newcommand{\re}[1]{Eq.~(\ref{#1})}
\newcommand{\ds}{\displaystyle}
\newcommand{\ov}[1]{\overline{#1}}
\newcommand{\hsp}{\hspace*{1pt}}

\begin{document}

\begin{center}
{\Large\bf
Nonequilibrium effects in hadronic fireball expansion}\\[5mm]

L.M.~Satarov$^{\,1,2}$, A.B.~Larionov$^{\,2,3}$, and I.N.~Mishustin$^{\,1,2}$
\end{center}

\begin{tabbing}
\hspace*{1.5cm}\=${}^1$\,\={\it Frankfurt Institute for Advanced Studies,~D--60438 Frankfurt~am~Main,~\mbox{Germany}}\\
\>${}^2$\>{\it National Research Center  ''Kurchatov Institute'', 123182~Moscow,~\mbox{Russia}}\\
\>${}^3$\>{\it Institut~f\"{u}r~Theoretische~Physik, Universit\"at Giessen,~D--35392 Giessen, Germany}
\end{tabbing}

\begin{abstract}
We consider {a spherical volume of hot and dense
hadronic matter (fireball) expanding} into a vacuum. It is assumed that initially the fireball
matter is in local thermal and chemical equilibrium with vanishing collective velocity.
The time evolution of the fireball is studied in parallel within the GiBUU transport
model and an ideal hydrodynamic model. The equation of state of an
ideal hadronic gas is used in the hydrodynamic calculation. The same set of
hadronic species is used in transport and fluid-dynamical simulations.
Initial coordinates and momenta of hadrons in transport simulations have been
randomly generated by using the Fermi and Bose distributions for (anti)baryons and mesons.
{The model results} for radial profiles of densities and collective velocities
of different hadronic species {are compared at different times}.
We find that two considered models predict {essential differences in time
evolution of hadron} abundances, which are especially pronounced
{for hyperonic species.}  This gives an evidence of
{a strong deviation} from chemical equilibrium in expanding hadronic matter.
\end{abstract}


\maketitle

\section{Introduction}

Physics of high energy heavy--ion collisions exhibits a rapid development during last several
decades. A lot of efforts has been made to extract the information on
properties of hot and dense
nuclear matter from experimental data. Numerous theoretical models have been proposed
to describe {the} complicated dynamics of multi-particle systems produced
in nuclear collisions. An important
place in understanding {the} main features of experimental data is still occupied by simple
fireball~\cite{Wes76,Bon78},
blast wave~\cite{Sie79,Sch93} and thermal~\cite{Bra95,Cle97} models.

More sophisticated hydrodynamic models (see e.g. a recent review~\cite{Cbm10})
have been very successful in reproducing a large amount of experimental data
in a wide range of bombarding energies. An attractive feature of these models
is their capability to study the sensitivity of observables to the equation
of state (EoS) of strongly interacting mater and, in particular, to a possible
deconfinement phase transition. The hydrodynamic models~(HDM) are usually applied
to describe the evolution of matter only at some intermediate stage of a heavy-ion collision.
They are not well suited for early and late stages since large deviations
from local thermodynamic equilibrium are expected in this case.

As a rule, these models assume the formation of a  {quasi--equilibrated fireball}
at an intermediate stage of the collision process.  {The} geometrical and
{thermodynamic} parameters of the initial fireball are normally
chosen to achieve the best fit of observed data.
The (2+1)--dimen\-sional HDM ~\cite{Bla87,Kol99,Per00,Bas00,Tea01} became especially popular in recent years.
These models describe the dynamics of a cylindrical fireball expanding in transverse
and longitudinal directions.

Of course, the {hydrodynamic} approach can not be directly applied to
simulate the late stages of a heavy-ion reaction when collisions of particles become too rare
to maintain the thermodynamic equilibrium.
The assumption of an instantaneous transition to the
collisionless propagation of particles (''freeze-out'') has been
introduced~\cite{Mil58,Coo74} to calculate the asymptotic particle spectra in HDM. However,
the direct kinetic
simulations of space--time distributions of last--collision points in relativistic nuclear
reactions show~\cite{Bra95a} that the freeze-out is in fact a continuous process,
even in central
interactions of heaviest nuclei (for details, see~\cite{Cbm10,Kno09}). It has been also
conjectured~\cite{Beb92,Cle99} that the stages of a ''chemical'' (saturation of particle
multiplicities) and ''kinetic''
(saturation of particle spectra) freeze-out should be separated in time.
It is believed that the chemical freeze-out takes place earlier because of much faster drop of inelastic
collision rates as compared to elastic ones in expanding
matter. On the other hand, there are several fluid--dynamical studies~\cite{Bro01,Pro09} where experimental
particle spectra have been successfully reproduced without introducing such a separation.

{Thermodynamic} equilibrium is not postulated in microscopic transport models of nuclear collisions.
Currently, the UrQMD~\cite{Bas98}, GiBUU~\cite{Gib09,Bus11}, HSD~\cite{Ehe96,Cas05}, QGSM~\cite{Ame93}
and AMPT~\cite{Lin01} transport models are widely used for simulating relativistic nuclear collisions.
These models are not well suited for describing high density states of
hadronic matter, where multiparticle (non-binary) channels of hadronic interactions are presumably
important. As a rule,
contributions of such channels are disregarded in existing versions of transport models. On the other
hand, as demonstrated in Refs.~\cite{Rap01,Gre01,Cas02}, multimesonic channels
of baryon--antibaryon pair production might be responsible for high yields of antiprotons and antihyperons
observed at SPS and RHIC energies.

A more realistic description of the freeze-out processes in heavy--ion collisions
can be achieved in a hybrid ''hydro--cascade'' model~\cite{Bas00,Tea01,Non07,Hir08,Pet08}.
In this approach, hydrodynamic and cascade simulations
are applied, respectively, for intermediate and late stages of the reaction.
The hybrid model implicitly assumes the existence of a space-time region
where the hydrodynamic and cascade simulations give similar results.
{The} characteristics
of hydrodynamic flow taken at a certain hypersurface are used for generating phase-space coordinates
of hadrons which serve as an input for subsequent transport simulations.

Although such a procedure seems to be intuitively justified, it should be verified for more or less
typical situations. In the present work, a spherical expansion of an ideal hadronic gas
is simulated in parallel by using
the transport and hydrodynamic models. Specifically we use a non-viscous  hydrodynamics with an
ideal--gas EoS and the transport GiBUU model~\cite{Gib09,Bus11}
(without mean--field interactions).
The initial fireball is assumed to be in thermodynamic equilibrium.
Thermodynamic parameters of this state are used to generate randomly the set of
initial hadrons
for subsequent transport calculations. By the direct comparison of results predicted
{by} the GiBUU and HDM
we investigate possible deviations from thermal and chemical equilibrium in the course of
expansion.

{Finally,} we would like to mention several earlier
works~{\cite{Bir83,Bar88,Gor92}}
which studied the particle composition in an expanding fireball {within a
''hadrochemical'' approach. There a set of rate equations for particle abundances
was combined with a simplified hydrodynamic description. But the calculations}
{had been performed} {for a spatially uniform background
neglecting surface effects.} The direct comparison between hydrodynamic and transport
calculations have been
recently made in the case of a cylindrical gluonic fireball in Ref.~\cite{Bou09}

The paper is organized as follows: parameters of the initial fireball are given in Sec.~IIA,
our hydrodynamic model is formulated in Sec.~IIB. In this section we also show radial profiles of
densities and collective velocities obtained by numerical solution of fluid-dynamical equations.
Our transport model is formulated in Sec.~IIC.
In Sec.~III we compare the predictions of the {hydrodynamic} and transport models.
Our conclusions are presented in Sec.~IV. The procedure of initial event generation is described in
Appendix.

\section{Description of the models}

\subsection{Initial state}~\label{sinit}


{We assume that at $t=0$} the fireball is locally equilibrated and has vanishing collective velocity.
The initial radial profiles of energy ($\varepsilon$)
and baryon~($n$) densities {are parametrized} by the Woods--Saxon distribution with some radius~$R$ and diffuseness $a$:
\bel{icon}
\varepsilon(r,0)=\varepsilon_0\hsp W(r),~~~n(r,0)=n_0\hsp W(r),~~W(r)=\left[\hsp\exp\left(\frac{r-R}{a}\right)+1\right]^{-1}.
\ee
Below we choose the following parameters:
\bel{icon1}
R=6~\textrm{fm}, a=0.3~\textrm{fm}, \varepsilon_0=1.3~\textrm{GeV/fm}^3, n_0=0.45~\textrm{fm}^{-3}.
\ee
The initial values of energy and baryon densities at the fireball center, $\varepsilon_0$ and $n_0$,
are typical for high density states of hadronic matter created in heavy--ion collisions
at $E_{\rm lab}\simeq 10$ AGeV (see e.g.~Ref.~\cite{Mer11}).

\subsection{Hydrodynamic simulation of fireball expansion}~\label{hyds}

In this section we describe the HDM for simulating the fireball expansion.
The model assumes that deviations from local equilibrium are small at any space-time point~$(\bm{r},t)$.
In~this approximation, the single-particle phase-space distribution function~(DF) of $i$-th hadronic species
is equal to a locally equilibrated DF $f^{(eq)}_i$ characterized by certain temperature $T$, chemical potential
$\mu_i$ and collective 3-velocity $\bm{v}$. Considering the hadronic
system  as a mixture of ideal gases of (anti)baryons and mesons, one can
approximate $f^{(eq)}_i$ by the Fermi or Bose DF ($\hbar=c=1$):
\bel{fbdf}
f^{\hsp (eq)}_i(\bm{r},\bm{p},t)=\frac{g_i}{(2\pi)^3}\left[\exp\left(\frac{\widetilde{E}_i-\mu_i}{T}\right)
\pm 1\right]^{-1}\,,
\ee
where $g_i$ is the spin-isospin degeneracy of $i$-th hadronic species~\footnote
{
We disregard the isospin and Coulomb effects.
},
$\widetilde{E}_i=\gamma\hsp (E_i-\bm{p}\bm{v})$
is a~single-particle energy in the {local rest frame (LRF)},
$E_i=\sqrt{m_i^2+\bm{p}^{\hspace*{0.5pt}2}}$ is
the corresponding energy in an arbitrary frame,
$m_i$ is the mass of $i$-th hadrons, and $\gamma=(1-\bm{v}^2)^{-1/2}$.
Plus or minus in the r.h.s. of \re{fbdf} correspond, respectively, to fermions or bosons.

In our fluid--dynamical calculations we assume a fireball matter
to be in chemical equilibrium with respect to strong interactions and decays of hadrons.
In this case, $\mu_i$ may be represented by linear
combinations of the baryon ($\mu$) and strange ($\mu_S$) chemical potentials:
\bel{hcep}
\mu_i=B_i\mu+S_i\mu_S\,,
\ee
where \mbox{$B_i=0,\pm 1$} and \mbox{$S_i=0, \pm 1, \pm 2\ldots$} are, respectively, the baryon
and strangeness quantum numbers of the $i$-th hadrons.

The values of the net baryon ($n$),  strangeness ($n_S$) and
energy ($\varepsilon$) densities {in LRF},
as well as pressure ($P$) may be expressed~\cite{Sat09} in terms of integrals of the DF (\ref{fbdf})
over the 3-momentum in {this} frame:
\bel{therq}
\left(\begin{array}{l}
n\\n_S\\\varepsilon\\P
\end{array}\hspace*{-2pt}\right)=\sum_i\int d^{\hsp 3}\tilde{p}\left(\begin{array}{l}
B_i\\S_i\\\widetilde{E}_{i}\\\frac{\ds\tilde{p}^{\,2}}
{\ds 3\widetilde{E}_{i}}\hspace*{-2pt}\end{array}\right) f^{\hsp (eq)}_i\,,
\ee
where the sum is taken over all hadronic species. Assuming that the strangeness density~\mbox{$n_S=0$} at
any space-time point, and using Eqs.~(\ref{fbdf})--(\ref{therq}), one can express all thermodynamic
quantities (e.g. $T, \mu, \mu_S, P$) in terms of two independent
variables $n$ and $\varepsilon$. In particular, in this way we obtain the
EoS of the fireball matter, $P=P(n,\varepsilon)$, which is used in our HDM (see below).

\renewcommand{\baselinestretch}{0.5}
\begin{table}
\caption{Hadronic species included in the calculation.
}
\label{tab1}
\footnotesize
\begin{ruledtabular}
\begin{tabular}{ccrcrcll|ccrcrcll}
hadron   & $m_i$ (GeV) &  $B_i$ & $S_i$ & $I_i$ & $g_i$ & $N_i$ & $N_{\ov{i}}$
& hadron & $m_i$ (GeV) &  $B_i$ & $S_i$ & $I_i$ & $g_i$ & $N_i$ & $N_{\ov{i}}$\\[1pt]
\hline
$\pi$             & 0.140  & 0 & 0    & 1   & 3  & 256.9 &      & $\Delta$\hsp(1600) &1.600 & 1 &  0   & 3/2 & 16 & 19.0 & 0.46  \\
$K$               & 0.496  & 0 & 1    & 1/2 & 2  & 94.0  &      & $\Delta$\hsp(1620) &1.630 & 1 &  0   & 3/2 & 8  & 8.2  & 0.20  \\
$\ov{K}$          & 0.496  & 0 & $-1$ & 1/2 & 2  & 28.4  &      & $N$\hsp(1650)      &1.655 & 1 &  0   & 1/2 & 4  & 3.6  & 0.089 \\
$\eta$            & 0.543  & 0 & 0    & 0   & 1  & 21.4  &      & $\Sigma$\hsp(1660) &1.660 & 1 & $-1$ &  1  & 6  & 2.9  & 0.24  \\
$\rho$            & 0.776  & 0 & 0    & 1   & 9  & 73.6  &      & $\Lambda$\hsp(1670)&1.670 & 1 & $-1$ &  0  & 2  & 0.93 & 0.076 \\
$\omega$          & 0.782  & 0 & 0    & 0   & 3  & 23.9  &      & $\Sigma$\hsp(1670) &1.670 & 1 & $-1$ &  1  & 12 & 5.6  & 0.50  \\
$\sigma$          & 0.800  & 0 & 0    & 0   & 1  & 7.4   &      & $\Omega^-$         &1.672 & 1 & $-3$ &  0  &  4 & 0.55 & 0.50  \\
$K^*$             & 0.892  & 0 & 1    & 1/2 & 6  & 53.9  &      & $N$\hsp(1675)      &1.675 & 1 &  0   & 1/2 & 12 & 8.9  & 0.24  \\
$\ov{K}^{\hsp *}$ & 0.892  & 0 & $-1$ & 1/2 & 6  & 16.3  &      & $N$\hsp(1680)      &1.685 & 1 &  0   & 1/2 & 12 & 9.4  & 0.23  \\
$N$               & 0.939  & 1 & 0    & 1/2 & 4  & 107.6 & 2.5  & $\Lambda$\hsp(1690)&1.690 & 1 & $-1$ &  0  &  4 & 1.7  & 0.14  \\
$\eta^{\hsp\prime}$& 0.958 & 0 & 0    & 0   & 1  & 3.7   &      & $N$\hsp(1700)      &1.700 & 1 &  0   & 1/2 &  8 & 5.8  & 0.14  \\
$\phi$            & 1.020  & 0 & 0    &  0  & 3  & 8.4   &      & $\Sigma$\hsp(1750) &1.750 & 1 & $-1$ &  1  &  6 & 1.9  & 0.15  \\
$\Lambda$         & 1.116  & 1 & $-1$ &  0  & 2  & 13.2  & 1.0  & $\Sigma$\hsp(1775) &1.775 & 1 & $-1$ &  1  & 18 & 5.0  & 0.41  \\
$\Sigma$          & 1.189  & 1 & $-1$ &  1  & 6  & 28.1  & 2.2  & $\Lambda$\hsp(1800)&1.800 & 1 & $-1$ &  0  &  2 & 0.49 & 0.040 \\
$\Delta$          & 1.232  & 1 &  0   & 3/2 & 16 & 110.6 & 2.6  & $\Lambda$\hsp(1810)&1.810 & 1 & $-1$ &  0  &  2 & 0.47 & 0.038 \\
$\Xi$             & 1.315  & 1 & $-2$ & 1/2 & 4  & 5.7   & 1.5  & $\Lambda$\hsp(1820)&1.820 & 1 & $-1$ &  0  &  6 & 1.3  & 0.11  \\
$\Sigma$\hsp(1385)& 1.385  & 1 & $-1$ &  1  & 12 & 22.1  & 1.8  & $\Lambda$\hsp(1830)&1.830 & 1 & $-1$ &  0  &  6 & 1.3  & 0.10  \\
$\Lambda$\hsp(1405)& 1.406& 1 & $-1$ &  0   & 2  & 3.3   & 0.27 & $N$\hsp(1880)      &1.880 & 1 &  0   & 1/2 &  8 & 2.4  & 0.059 \\
$N$\hsp(1440)     & 1.440  & 1 &  0   & 1/2 & 4  & 10.3  & 0.25 & $\Lambda$\hsp(1890)&1.890 & 1 & $-1$ &  0  &  4 & 0.63 & 0.052 \\
$\Lambda$\hsp(1520)&1.520  & 1 & $-1$ &  0  & 4  & 3.8   & 0.31 & $\Delta$\hsp(1905) &1.890 & 1 &  0   & 3/2 & 24 & 6.9  & 0.17  \\
$N$\hsp(1520)     & 1.520  & 1 &  0   & 1/2 & 8  & 14.0  & 0.34 & $\Delta$\hsp(1910) &1.910 & 1 &  0   & 3/2 &  8 & 2.1  & 0.051 \\
$\Xi$\hsp(1530)   & 1.533  & 1 & $-2$ & 1/2 & 8  & 4.0   & 1.1  & $\Sigma$\hsp(1915) &1.915 & 1 & $-1$ &  1  & 18 & 2.5  & 0.21  \\
$N$\hsp(1535)     & 1.530  & 1 &  0   & 1/2 & 4  & 6.7   & 1.6  & $\Delta$\hsp(1930) &1.960 & 1 &  0   & 3/2 & 24 & 4.8  & 0.12  \\
$\Lambda$\hsp(1600)&1.600  & 1 & $-1$ &  0  & 2  & 1.3   & 0.11 & $\Delta$\hsp(1950) &1.930 & 1 &  0   & 3/2 & 32 & 7.5  & 0.19  \\
\end{tabular}
\end{ruledtabular}
\normalsize
\end{table}
\renewcommand{\baselinestretch}{1.2}
To facilitate the comparison with transport simulations,
in our hydrodynamic calculations we use the EoS with the same set of
hadrons and hadronic resonances as in the GiBUU model
(see Appendix~A of Ref.~\cite{Bus11}). The set of nonstrange baryons ($B_i=1, S_i=0$) includes nucleons ($N$), isobars ($\Delta (1232)$)
and their excited states. We also take into account {hyperons stable with respect to strong decays}
($Y=\Lambda,\Sigma,\Xi,\Omega^-$) and hyperonic resonances ($Y^*=\Sigma (1385), \Lambda (1405),\ldots$).
In addition to baryons we include corresponding antibaryons ($B_i=-1$).
The mesonic set ($B_i=0$) consists of stable mesons $\pi,K,\ov{K},\eta,\eta^{\hsp\prime}$ and resonances
$\rho,\omega,\sigma,\varphi,K^*,\ov{K}^{\hsp *}$.

The list of baryons and mesons included in our calculations is given in Table~\ref{tab1}~\footnote
{
The columns labeled by $I_i$ show isospin quantum numbers of corresponding hadrons.
}.
The thermodynamic properties of a fireball matter, in particular,
its EoS, are obtained assuming zero widths of resonances.
Using Eqs.~(\ref{icon})--(\ref{therq})
and characteristics of hadrons from Table~\ref{tab1} we calculate the initial radial
profiles of temperature $T$,
chemical potentials~$\mu,\mu_S$ and partial densities of hadronic species,
\mbox{$n_i=\int d^{\hsp 3}\widetilde{p}\hsp f^{\hsp (eq)}_i$}.
This calculation gives the values \mbox{$T\simeq 180~\textrm{MeV}, \mu\simeq 328~\textrm{MeV}$},
\mbox{$\mu_S\simeq 110~\textrm{MeV}$} at the fireball center~(\mbox{$r=0$}).
The last two columns of Table~\ref{tab1} show the equilibrium
multiplicities of $i$-th hadrons~($N_i$) and antibaryons ($N_{\ov{i}}$) in the initial state.
These quantities
are obtained by the volume integration of the corresponding
partial densities. One can see that most
abundant hadrons at $t=0$  are pions, nucleons and isobars. The total multiplicity of initial
antibaryons (approximately 18.6)
is much smaller than the total baryon number of the fireball
\mbox{$B_{\hsp\rm tot}=\sum\limits_i |B_i| (N_i-N_{\ov{i}})=\int d^{\hsp 3}r n\hsp\gamma\simeq 417$}\,.

To describe the dynamics of spherical fireball expansion, we solve the equations of
the \mbox{(1+1)--dimensional} ideal hydrodynamics. They can be written in the form~\cite{Lan87}:
\begin{eqnarray}
&&\frac{\partial {\cal N}}{\partial\hsp t}+\left(\frac{\partial}
{\partial\hsp r}+\frac{2}{r}\right) (v{\cal N}) = 0\,,\label{hydn}\\
&&\frac{\partial E}{\partial\hsp t}+\left(\frac{\partial}
{\partial\hsp r}+\frac{2}{r}\right)M = 0\,,\label{hyde}\\
&&\frac{\partial M}{\partial\hsp t}+\left(\frac{\partial}
{\partial\hsp r}+\frac{2}{r}\right)(vM+P) = 0\,.\label{hydm}
\end{eqnarray}
Here $v=\bm{v}\bm{r}/r$ is the radial component of fluid velocity, ${\cal N}=n\hsp\gamma$
is the net baryon density in the {fireball c.m.}
frame,
$E$ and $M$ are the components of the energy--momentum tensor:
\bel{emts}
E=T^{\hsp 00}=\gamma^2(\varepsilon+v^2P),~~M=T^{\hsp 0r}=v\hsp (E+P)\,.
\ee
The initial conditions for these equations are given by the relations (\ref{icon}) and \mbox{$v(r,0)=0$}.
Solving (\ref{hydn})--(\ref{emts}) together with the EoS \mbox{$P=P\hsp (n,\varepsilon)$} gives
the radial profiles of $n,\varepsilon, v$ at fixed~\mbox{$t>0$}. Similar hydrodynamic studies of
a spherical fireball expansion by using
simplified EoS\hspace*{1pt}s have been performed earlier in Refs.~\cite{Mis83,Bis95,Ris96}.

The numerical solution of Eqs.~(\ref{hydn})--(\ref{hydm}) has been obtained by using the
flux-corrected transport algorithm
SHASTA~\cite{Bor73,Ris95}. Typically, we choose the cell sizes \mbox{$\Delta\hsp r=2.5\cdot 10^{-2}~\textrm{fm}$},
\mbox{$\Delta\hsp t=5\cdot 10^{-3}~\textrm{fm/c}$} of the \mbox{$(r,t)$} grid. Similarly to Ref.~\cite{Mer11}, in this calculation
we use linear interpolations of the $P\hsp (n,\varepsilon)$ table prepared with fixed steps in $n$ and $\varepsilon$. We have checked
that our numerical scheme conserves the total baryon number, energy and entropy of the fireball with relative accuracy better that 1\%.

     \begin{figure*}[htb!]
         \vspace*{3mm}
          \centerline{\includegraphics[width=\textwidth]{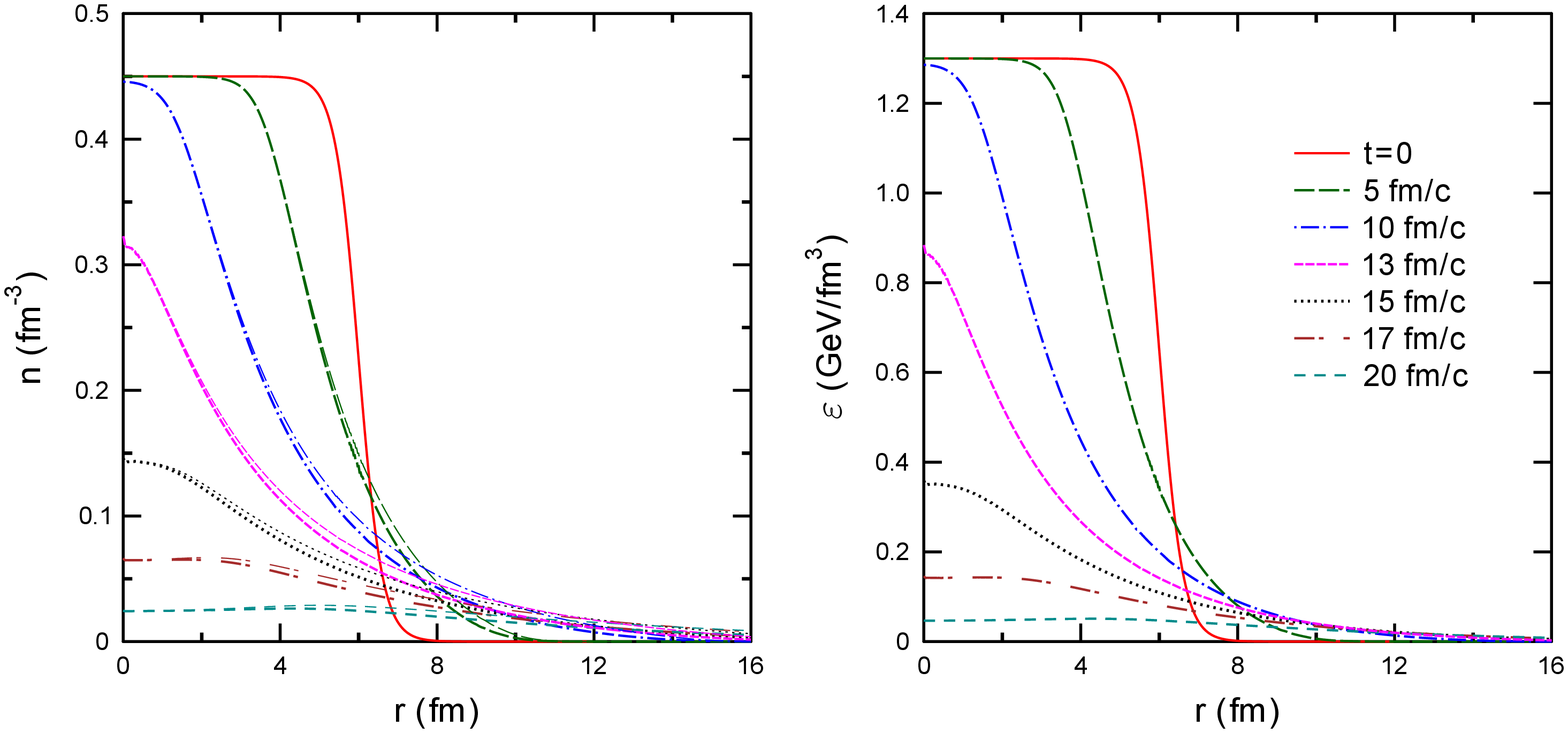}}
        \caption{
        {The} radial profiles of the baryon density (left panel)
        and energy density (right panel) calculated within HDM.
        }
        \label{fig1}
        \end{figure*}
\begin{figure*}[htb!]
          \vspace*{3mm}
          \centerline{\includegraphics[width=\textwidth]{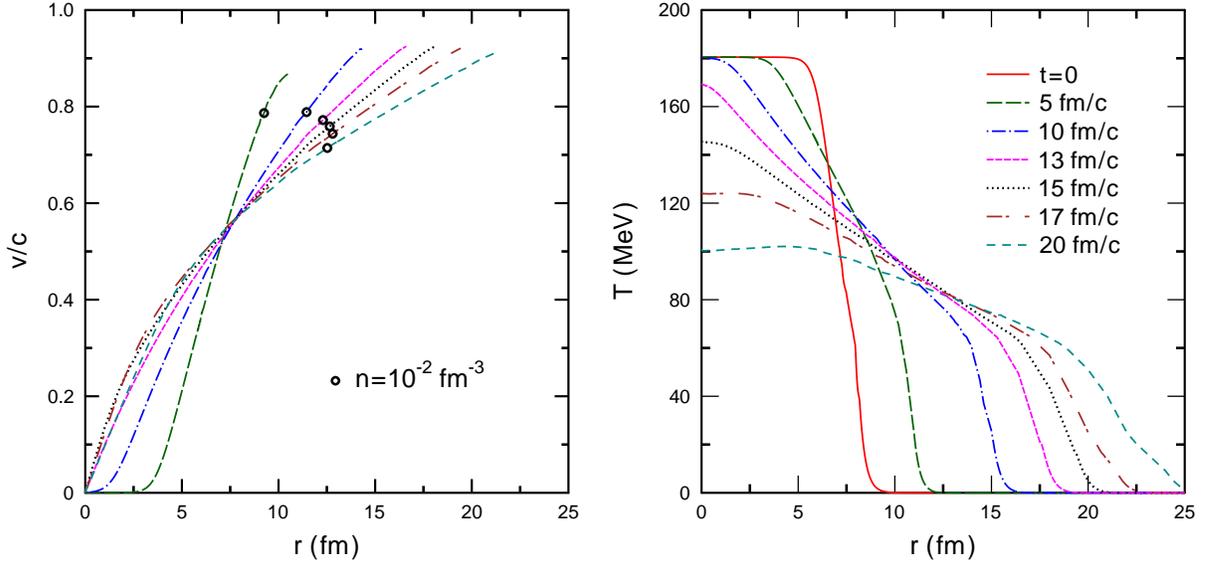}}
        \caption{
        The radial profiles of the collective velocity (left panel) and temperature
        (right panel) calculated in HDM.
        {The} outer parts of velocity profiles corresponding to densities $n<10^{-3}\,\textrm{fm}^{-3}$
        are omitted. {The} dots in the left panel correspond to spatial points where $n=10^{-2}\,\textrm{fm}^{-3}$.
        }
        \label{fig2}
        \end{figure*}

The radial profiles of fluid-dynamical quantities calculated for different
times $t$ within the HDM are shown in Figs.~\ref{fig1}--\ref{fig2}.
{In particular, thick and thin lines in the left panel of Fig.~\ref{fig1}
show the profiles of $n$ and $\cal N$\hspace*{-1pt}, respectively.}
As one can see in Fig.~\ref{fig1}, a rare\-faction wave propagates from the
fireball periphery and reaches the fireball center at $t\sim 10~\textrm{fm/c}$\hsp.
At later times the matter in the whole fireball is involved in the expansion. At such
times the radial profile of collective velocity can be approximated as $v=H r^\alpha$ with $\alpha<1$.
In Fig.~\ref{fig2}\hsp , we do not show external parts of velocity
profiles corresponding to dilute regions of matter with~$n<10^{-3}\,\textrm{fm}^{-3}$.
These parts certainly can not be realistically described within the ideal
HDM because of large local Knudsen numbers.

\vspace*{2mm}
\subsection{The GiBUU transport model}~\label{gbtm}

To study possible deviations from local equilibrium, the expansion of the
hadronic fireball  has been also simulated within the GiBUU transport
model {(version 1.2.2)}~\cite{Gib09,Bus11,Lar10}.
The detailed description of the model is given in Ref.~\cite{Bus11}, below we
present only its short summary. 
We apply GiBUU in a cascade mode, i.e.~omitting contributions of nuclear and Coulomb potentials.
The evolution of hadronic phase-space DFs $f_i(\bm{r},\bm{p},t)$
due to free propagation of hadrons, as well as due to their two-body collisions
and resonance decays is described by kinetic equations with corresponding collision
terms. They are solved by using the standard test-particle method~\cite{BDG88}
which includes the Hamiltonian equations of motions for test particles
and their two-body collisions (or decays) generated by a Monte Carlo method.
In present calculations we use one test particle per hadron and consider a large
number of parallel ensembles which we call ''events''. Each event is characterized by
a specific sets of hadronic species, as well as their coor\-dinates~$\bm{r}_j(t)$ and
momenta $\bm{p}_j(t)$.

{The DF of $i$-th hadronic species is represented as follows}
\vspace*{2mm}
\bel{gidf}
f_i(\bm{r},\bm{p},t)=\left<\sum\limits_{j=1}^{N_i}\delta\left[\bm{r}-\bm{r}_j(t)\right]\hsp\delta
\left[\bm{p}-\bm{p}_j(t)\right]\right>,
\ee
\vspace*{2mm}
where the angular brackets denote averaging over all events
and the sum runs over all hadrons
of the type $i$ existing at given time $t$ in a given event~\footnote
{
Note, that, in general, the multiplicity of $i$-th hadrons, $N_i$, is a function of $t$\hsp.
}.
The hadrons propagate along straight-line trajectories
(\mbox{$\dot{\bm{r}}_j=\bm{p}_j/E_j,~\dot{\bm{p}}_j=0$})
between their two-body scatterings. The latter change the momenta abruptly and also
lead to production of new hadrons~\footnote
{
The geometrical criterium, suggested in Ref.~\cite{BDG88}, is used for generating
two--body scatterings.
}.
Resonances are allowed to decay during time evolution.


For calculating collective velocities of particles, one should
know radial profiles of the 4-current density
$J^{\mu}_i$ and the energy-momentum tensor $T^{\mu\nu}_i$ of the hadronic species~$i$.
As in the HDM, we consider initial states
with spherically symmetric distributions of particles.
Let us consider a thin shell occupying a region of points with radii between $r$
and \mbox{$r+\Delta r$}.
In the spherical coordinate system with the origin at the fireball center,
only the components of $J_i^\mu, T^{\mu\nu}_i$
with \mbox{$\mu,\nu=0,r$} are nonzero {in the limit of a large number of events}.
By using~\re{gidf} one can get the relations:
\vspace*{2mm}
\begin{eqnarray}
&&J_i^{\hspace*{0.5pt}0}(r,t)=\int d^{\hsp 3}p\hsp f_i=
\left<\sum\limits_{j=1}^{N_i}\Theta_j\right>,\label{hcur}\\
&&T_i^{\hsp 00}(r,t)=\int d^{\hsp 3}p\hsp E_i\hsp f_i=
\left<\sum\limits_{j=1}^{N_i}E_j\Theta_j\right>,\label{hede}
\end{eqnarray}
where \mbox{$\Theta_j=(4\pi\hspace*{0.5pt}r^2\Delta\hsp r)^{-1}$} if the radial
coordinate $r_j$ of hadron $j$ in a given event
falls into the interval $(r,r+\Delta\hsp r)$ and \mbox{$\Theta_j=0$} in the opposite case.
The second equalities in~(\ref{hcur})--(\ref{hede})
are obtained in the limit of small $\Delta\hsp r$ and large {number of events}. A similar
expression for $J_i^{\hspace*{0.5pt}r}$ takes place
after replacing $\Theta_j\to v_j^{\hsp r}\hsp\Theta_j$ in (\ref{hcur}), where
$v_j^{\hsp r}$ is the radial component of  the $j$-th hadron velocity.
The replacements of $\Theta_j$ by $v_j^{\hsp r}\hsp\Theta_j$ and $(v_j^{\hsp r})^2\hsp\Theta_j$
in (\ref{hede}) give, respectively, the corresponding
rela\-tions for~$T^{\hsp 0\hspace*{0.5pt}r}_i$ and $T^{\hsp rr}_i$.

For generating initial events we use a procedure similar to that suggested in Ref.~\cite{Pet08}.
It is described in Appendix. These events have been generated
{according to the Fermi and Bose DF of hadrons}~(\ref{fbdf}).
In principle, in this way one should obtain the same ensemble averaged initial distributions
in both models~\footnote
{
Note that using the quantum Fermi and Bose distributions instead of the corresponding
Boltzmann~DF for initial hadrons within GiBUU is not
fully consistent. Indeed, the Pauli blocking {is normally
included for nucleons in this model,} {but it is neglected in our present calculation
to reduce the CPU time} {(one should have in mind that this blocking should be negligible
at high temperatures considered here). On the other hand, the Bose--enhancement effects are completely
disregarded in GiBUU}.
}.
Unless otherwise stated, the GiBUU simulations {are averaged over 5000 events}.
Note that in our generating procedure
we do not fix the total strangeness and charge of particles in a single event. However, we have
checked that fluctuations of these quantities are relatively
small for typical events. In particular, absolute values of net total strangeness do not
exceed 1\% of $B_{\hsp\rm tot}$.

\section{Results}

\subsection{Comparison of density profiles}

\begin{figure*}[htb!]
     \vspace*{3mm}
          \centerline{\includegraphics[width=\textwidth]{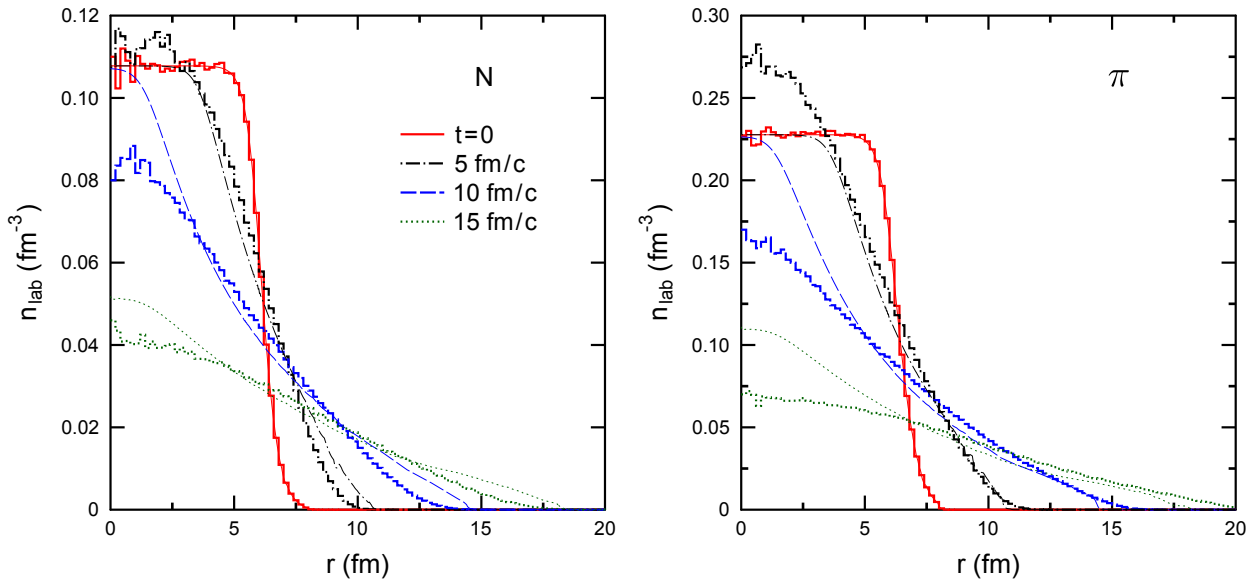}}
        \caption{
        The radial profiles of nucleon (left panel) and pion (right panel) densities calculated
        within the GiBUU (histograms) and hydrodynamic (thin lines) models.
        }
        \label{fig3}
        \end{figure*}

The process of fireball expansion into a vacuum has been simulated in parallel
within the hydrodynamic and GiBUU models.
In this section we compare the density profiles for different hadronic species.
Figure~\ref{fig3} shows the nucleon and pion densities in the {fireball c.m.}
frame. These densities are summed over different isospin states of corresponding hadrons.
Histograms in Fig.~\ref{fig3} are obtained within the GiBUU {model} by
using~\re{hcur} for \mbox{$i=N,\pi$}. Thin lines represent
{the profiles of densities}~$n_i\gamma$ obtained within the~HDM.
In the case of pions, deviations between two calculations are visible already
at relatively short times \mbox{$t\lesssim 5~\textrm{fm}/c$}\hsp .
In particular, one can clearly see an excess of pions predicted by GiBUU in a central region.
This effect can be explained by multi-hadron absorption
processes {missed in GiBUU} but implicitly included in the HDM.
Another possible reason is the Bose--enhancement effects neglected
in transport calculations{, but taken into account in generating initial events}
(see the footnote at the end of Sec.~\ref{gbtm}).
At later times the HDM densities are systematically larger than
the GiBUU predictions in this region.

\begin{figure*}[htb!]
          \centerline{\includegraphics[width=\textwidth]{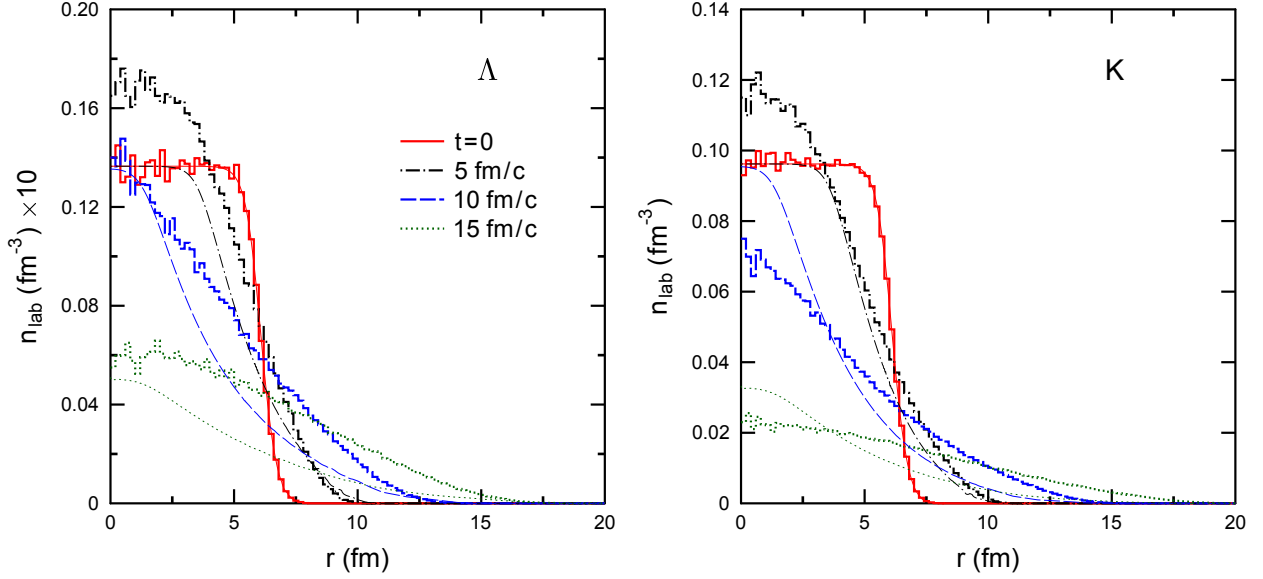}}
        \caption{
        Same as Fig.~\ref{fig3}, but for radial profiles of $\Lambda$ hyperons (left panel) and kaons (right panel).
        }
        \label{fig4}
        \end{figure*}

The density profiles of $\Lambda$ hyperons and kaons are shown in Fig.~\ref{fig4}.
One can see again that in contrast to the HDM, the particle densities
in the central region increase in GiBUU at early times.  However,
at $t\gtrsim 10~\textrm{fm}/c$\hsp , the local densities of $\Lambda$'s predicted by HDM decrease much faster
with time than those obtained in GiBUU. This is a consequence of {the} rapid drop
of {the} activation factor $\exp{[(\mu_\Lambda-m_\Lambda)/T]}$ {as the matter cools down}.
This factor is especially important for massive particles like hyperons and baryon resonances.
As we shall see in Sect.~\ref{pmult} a faster drop of densities gives rise to a
qualitatively different time evolution of heavy particles' multiplicities in the
HDM and GiBUU.

\subsection{Collective  velocities}
There is no unique way to determine {a} collective velocity (CV) of particles in
transport models. One possibility is to define this velocity (we denote it by
$\bm{v}_1$) as the velocity of {LRF} where the 3--vector of the current density vanishes
{in average over many events}.
In the case of spherically symmetric expansion{,} the CV has only the~radial component.
In general, within GiBUU, the hadrons of different types have different~CVs.
Using the Lorentz transformation to the frame where the radial current of $i$-th hadronic species equals
to zero, one {obtains} the relations:
\bel{colv1}
\widetilde{J}^r_i=0\to v_1^{(i)}=\frac{J^r_i}{J^{\hspace*{0.5pt}0}_i}\,.
\ee
The current density components in second equation are calculated by using (\ref{hcur}).

Another possibility is to find the velocity of reference frame with vanishing \mbox{3--momentum} flux.
We denote the corresponding CV by $\bm{v}_2$. From  the Lorentz transformation for
the energy-momentum tensor one has
\bel{colv2}
\widetilde{T}_i^{\hsp 0r}=0\to v_2^{(i)}=\frac{2\hsp T^{\hsp 0r}_i}
{T^{\hsp 00}_i+T^{rr}_i+\sqrt{(T^{\hsp 00}_i+T^{rr}_i)^2-
4\hsp (T^{\hsp 0r}_i)^2}}\,.
\ee
It can be shown that in the case of a local equilibrium both definitions give the
same values of~CV: \mbox{$v_1^{(i)}=v_2^{(i)}=v$},
where $v$ is the flow velocity in the HDM. Our GiBUU calculations
show that $v_1^{(i)}$ and $v_2^{(i)}$
are nearly equal for all kinds of hadrons. In the following we determine the~CV by using the
definition (\ref{colv1}).

    \begin{figure*}[htb!]
          \centerline{\includegraphics[width=\textwidth]{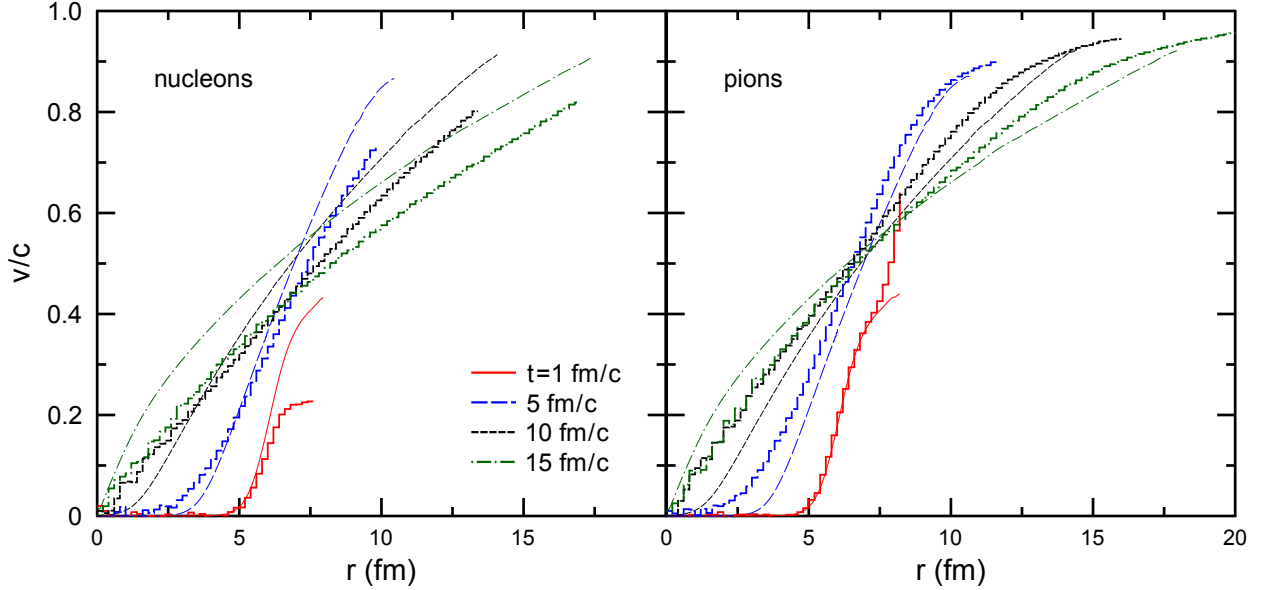}}
        \caption{Collective velocities of nucleons (left panel) and pions (right panel)
        as functions of radius~$r$ at different times indicated in the left panel.
        Thick and thin lines are calculated within the GiBUU and HDM, respectively.
        Outer parts of profiles corresponding to densities below $10^{-3}\,\textrm{fm}^{-3}$
        are omitted.
        }
        \label{fig5}
        \end{figure*}
The radial profiles of CV for nucleons and pions are shown in Fig.~\ref{fig5}
for several times~$t$. One can see that these CV are not equal to each other
in the transport calculation. At small~$t$ significant differences between $v^{(N)}$, $v^{(\pi)}$
and the hydrodynamic velocity exist only at the fireball periphery. At later times
the differences become visible for all $r$. The deviations between the models are
especially large for nucleons at~\mbox{$t\gtrsim 10$ fm/c}. These deviations
can be explained by dissipative effects effectively included in GiBUU,
but disregarded in the~HDM.

\subsection{Hadron abundances}\label{pmult}

In this section we compare evolutions of particle multiplicities predicted by the GiBUU
and~HDM. These multiplicities are obtained from radial profiles of partial densities by integration
over the fireball volume. In addition to the partial densities of ''free'' hadrons{,}~$n_i${,}
below we also calculate the corresponding total densities, $n^{\rm tot}_i$, which include hadrons
''hidden'' in heavier hadronic resonances~\cite{Beb92}:
\bel{nhid}
n^{\rm tot}_i=n_i+\sum_j d^{\hsp i}_j n_j\hsp,
\ee
where $d^{\hsp i}_j$ is {the} average number of $i$-th hadrons,
produced in decays \mbox{$j\to i+X$}. The sum in the r.h.s. runs over all resonances
included in our simulations (see Table~\ref{tab1}) and
having strong decays into $i$-th hadrons. The coefficients $d^{\hsp i}_j$
are calculated using branching ratios given in Ref.~\cite{Nak10}.
     \begin{figure*}[htb!]
          \vspace*{2mm}
          \centerline{\includegraphics[width=0.6\textwidth]{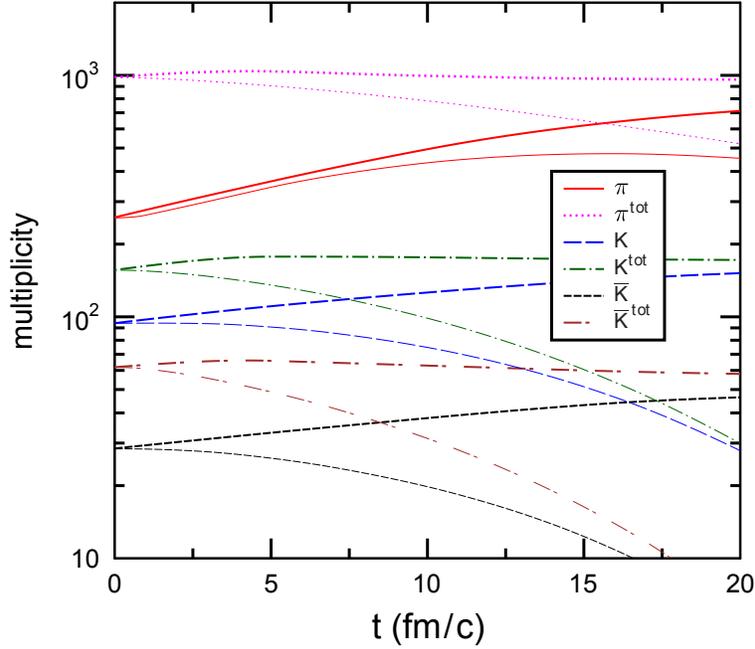}}
        \caption{
        Multiplicities of pions, kaons and antikaons
        as functions of time in the hadronic fireball expanding {into} vacuum.
        Thick (thin) lines are calculated within the GiBUU
        (hydrodynamic) model. $\pi^{\hsp\rm tot}, K^{\hsp\rm tot}$ and $\overline{K}^{\hsp\rm tot}$
        are total numbers of pions, kaons and antikaons including mesons,
        hidden in hadronic resonances (see~\re{nhid}).
        }
        \label{fig6}
        \end{figure*}
     \begin{figure*}[htb!]
          \centerline{\includegraphics[width=0.6\textwidth]{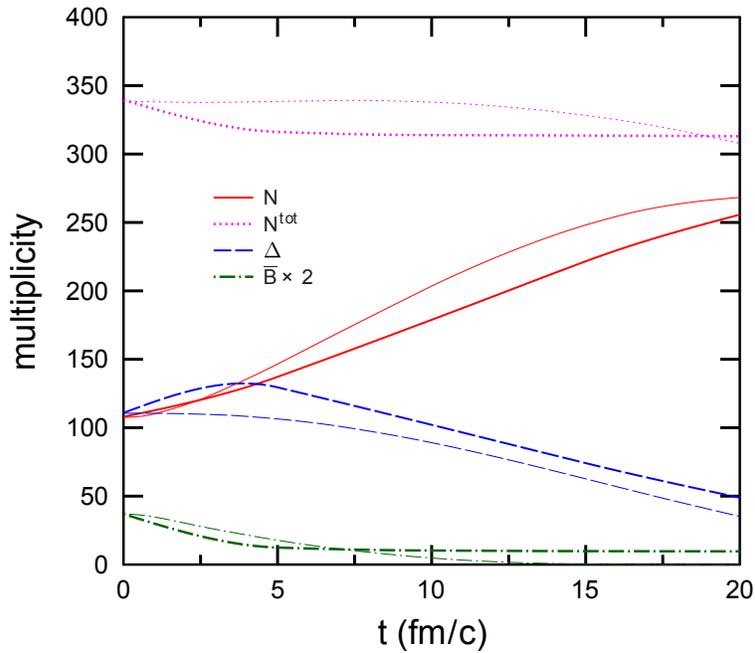}}
        \caption{
        Same as Fig.~\ref{fig6}, but for nucleons, $\Delta$--isobars and antibaryons.
        $N^{\rm tot}$ is {the} total number of nucleons including those,
        hidden in hadronic resonances.
        }
        \label{fig7}
        \end{figure*}
     \begin{figure*}[htb!]
          \centerline{\includegraphics[width=0.6\textwidth]{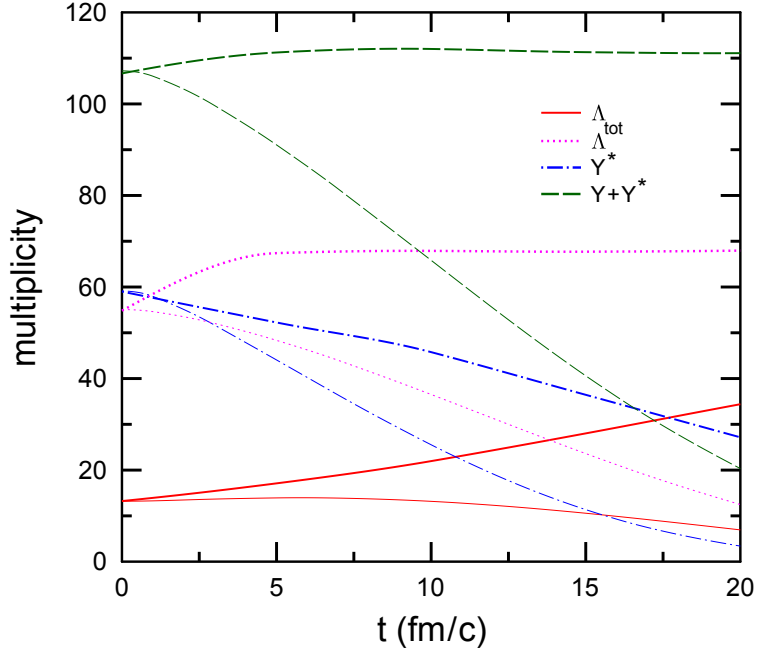}}
        \caption{
        Same as Fig.~\ref{fig7}, but for multiplicities of hyperons. $Y (Y^*)$
        denote multiplicities of all stable (unstable) hyperons.
        $\Lambda^{\rm tot}$ is {the} total number of $\Lambda$'s including those,
        hidden in hyperonic resonances.
        }
        \label{fig8}
        \end{figure*}
     \begin{figure*}[tbh!]
          \centerline{\includegraphics[width=0.6\textwidth]{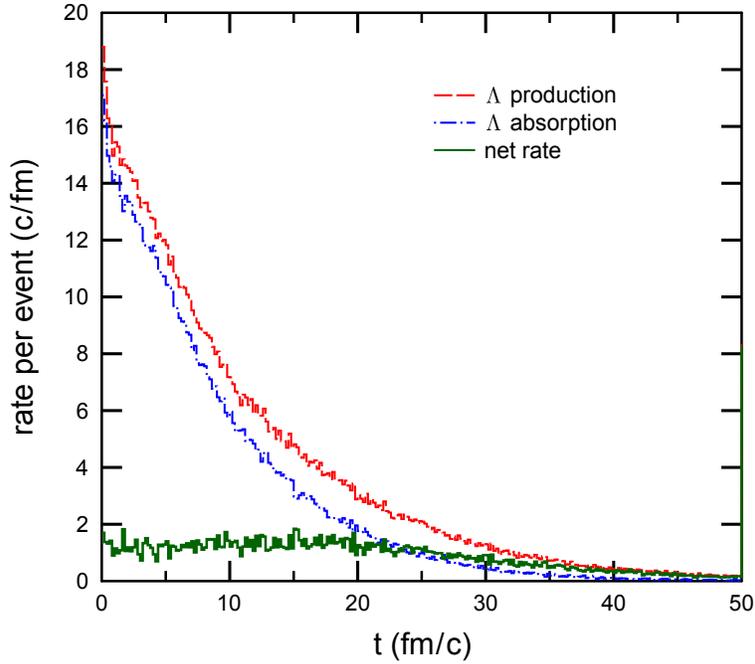}}
        \caption{
        Total production (dashed line) and absorption (dash-dotted line) rates of $\Lambda$ hyperons as functions
        of time in GiBUU. The solid line shows the net (production minus absorbtion) rate.
        }
        \label{fig9}
        \end{figure*}

Figures~\ref{fig6}--\ref{fig8} show the time evolution of hadron multipli\-cities
pre\-dic\-ted by the GiBUU and hydrodynamic models. One can see a qualitative difference between
the results of two calculations. In particular, the GiBUU model predicts nearly
constant total multiplicities of pions, kaons and hyperons including hadrons
hidden in resonances. However, these quantities decrease noticeably within the HDM.

The two models predict very different time evolution of multiplicities of
hyperons and hyperonic resonances. As one can see from
Fig.~\ref{fig8}, the multiplicity of $\Lambda$ hyperons noticeably
increases in GiBUU, but decreases in the~HDM\hsp.
According to our transport calculation, the number of $\Lambda$'s increases from~13 at $t=0$
to about 50 at $t=40~\textrm{fm}/c$. Within this time interval many
$\Lambda$ particles are still ''bound'' in hyperonic resonances. When time
increases from $0$ to $40~\textrm{fm}/c$, the relative fraction of bound $\Lambda$'s decreases
from about 80\% to 30\%~\footnote
{
{In calculating  $\Lambda^{\rm tot}$ we include $\Lambda$ hyperons
from the $\Sigma^{\hsp 0}\to\Lambda\gamma$ decays. Without this contribution, the relative fraction
of bound $\Lambda$'s changes from 70\% to 7\% in the same time interval.}
}.
A long duration of~$\Lambda$ production in GiBUU is
directly connected with a slow decrease of {the} abundances
of hyperonic resonances (compare dash--dotted curves in Fig.~\ref{fig8}).
This in turn follows from significant regeneration
of these resonances in baryon--baryon and baryon-meson collisions.
     \begin{figure*}[htb!]
          \centerline{\includegraphics[width=0.6\textwidth]{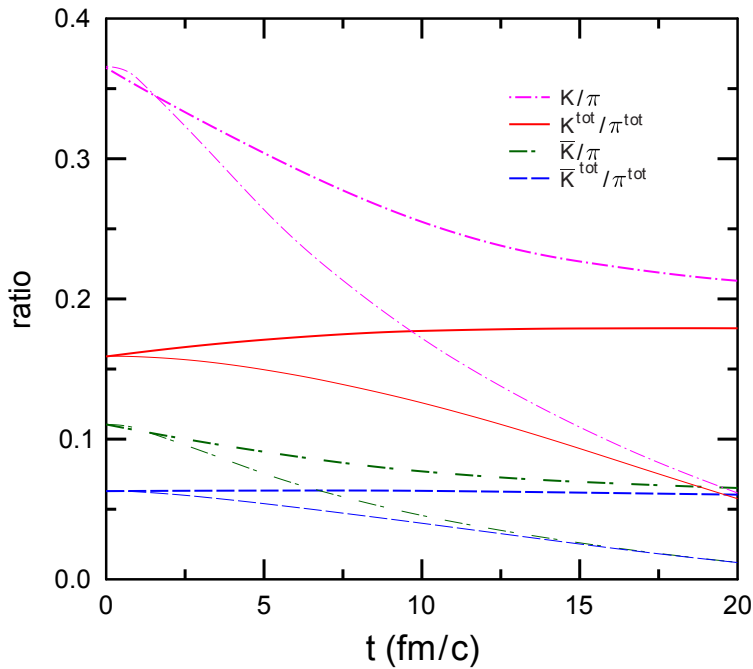}}
        \caption{
        (Anti)kaon to pion ratios as functions of time in expanding fireball.
        Thick (thin) lines are calculated within the GiBUU
        (hydrodynamic) model.
        }
        \label{fig10}
        \end{figure*}

To analy{z}e relative importance of different channels
of $\Lambda$ production and absorption we have
calculated total rates of such processes in GiBUU. This analysis shows that for our
parameters of initial state, the most important production channels
are decays $Y^*\to\Lambda M$ (here~$M$ denotes a nonstrange meson) and \mbox{$\Sigma B\to\Lambda X$},
\mbox{$Y^*B\to\Lambda X$} reactions ($B$ is a nonstrange baryon). On the other hand, most significant
channels of $\Lambda$ absorption are the reactions \mbox{$\Lambda M\to Y^*X$} and
\mbox{$\Lambda B\to (\Sigma\hsp,Y^*)\hsp X$}. As one can see in Fig.~\ref{fig9}, both
the $\Lambda$ production and absorption rates are rather high at $t\lesssim 20~\textrm{fm}/c$, but they nearly
compensate each other: the net production rate is of the order of $1~c/\text{fm}$ at
$t\lesssim 40~\textrm{fm}/c$. Only later the multiplicity of $\Lambda$ hyperons saturates.
It is interesting that large characteristic times (exceeding approximately
$40~\textrm{fm}/c$) of $\Lambda$ production have been obtained earlier in the UrQMD~\cite{Bas00}
and BUU~\cite{Bra00} calculations.

Figure~\ref{fig10} shows the time dependence of (anti)kaon to pion multiplicity ratios. Again, one can see
significant differences between the model predictions. For example,
the ratios $K^{\rm tot}/\pi^{\rm tot}$ and $\ov{K}^{\rm tot}/\pi^{\rm tot}$
are practically constant in the GiBUU calculation. However, they rapidly
decrease with time in the HDM. These results may imply large deviations from chemical equilibrium in
expanding baryon-rich hadronic matter created at FAIR, NICA and lowest RHIC energies.


\subsection{Evolution of energy distributions}

    \begin{figure*}[htb!]
          \centerline{\includegraphics[width=\textwidth]{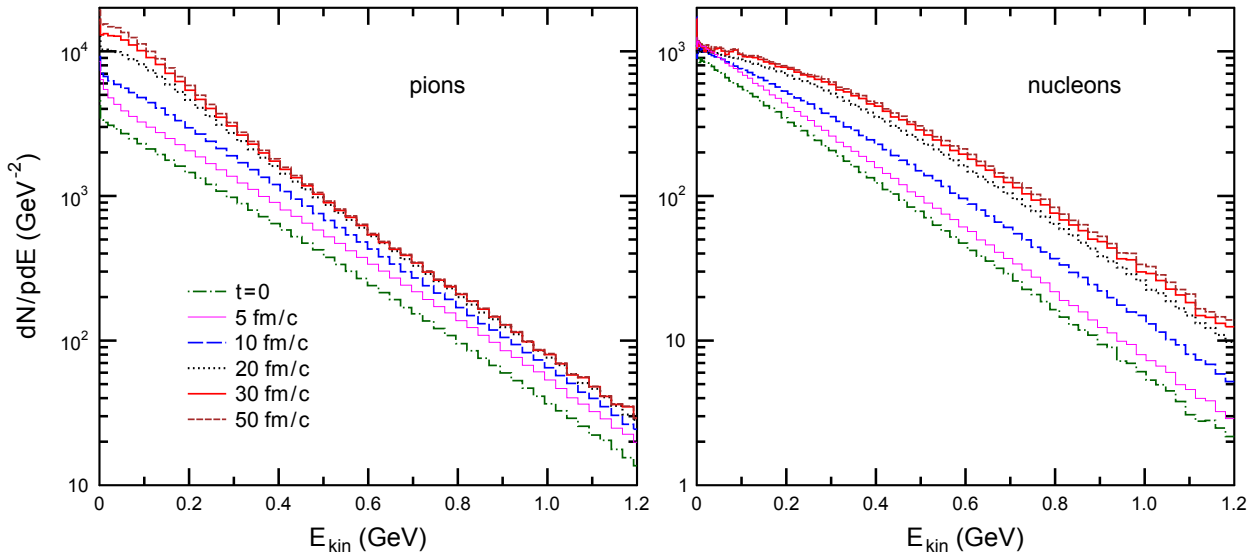}}
        \caption{
        Kinetic energy distributions of pions (left panel) and nucleons (right panel) in expanding fireball.
        Lines show the GiBUU results at different time moments $t$.
        }
        \label{fig11}
        \end{figure*}
\begin{figure*}[htb!]
          \vspace*{2mm}
          \centerline{\includegraphics[width=\textwidth]{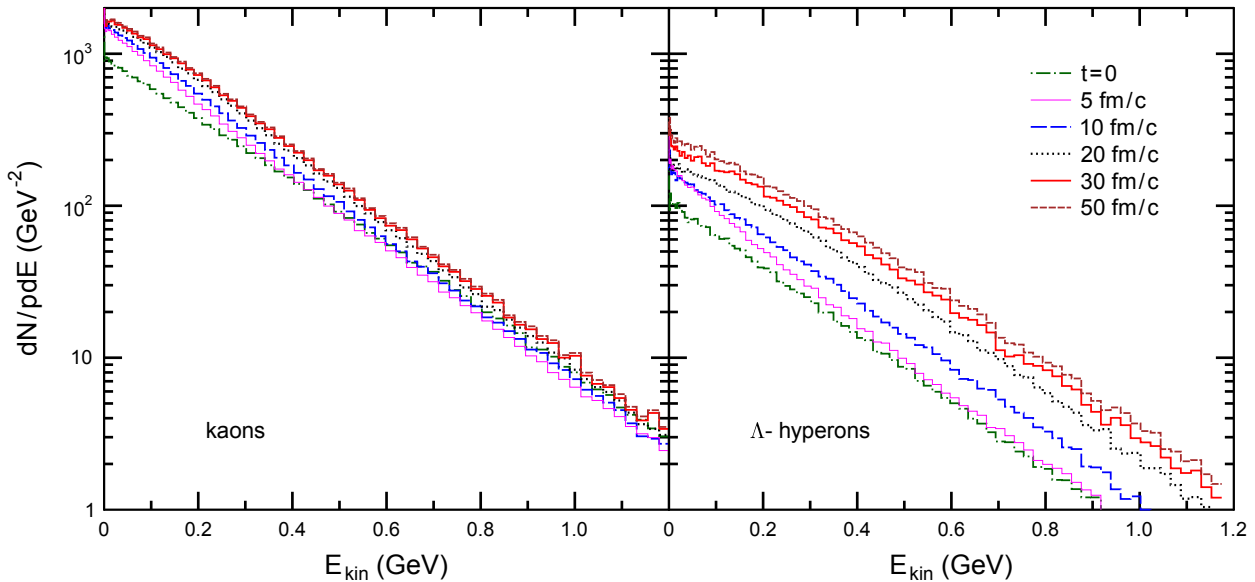}}
        \caption{
        Same as Fig.~{\ref{fig11}}, but for kaons (left panel) and $\Lambda$'s (right panel).
        }
        \label{fig12}
        \end{figure*}

It is instructive to see, how particle momentum spectra evolve with time in GiBUU.
In Figs.~\ref{fig11}--\ref{fig12}{\hsp ,} we show the kinetic energy
distributions of pions, nucleons, kaons and~\mbox{$\Lambda$ hyperons}
at different~$t$. One can clearly see the importance of final state rescatterings and
resonance decays in formation
of these spectra. The collective expansion results in concave and convex shapes of
momentum distributions for pions
and nucleons, respectively. These effects are less pronounced for kaons and $\Lambda$'s.
The GiBUU simulations predict the sequential behavior of freeze-out.
Indeed, the formation of asymptotic distributions is taking place
at different times for different hadrons.
First, this happens for kaons (at $t\sim 20~\textrm{fm}/c$), then for pions and nucleons
($t\gtrsim 30~\textrm{fm}/c$) and finally, for $\Lambda$ hyperons ($t\gtrsim 50~\textrm{fm}/c$).
One can not clearly distinguish the stages of ''chemical'' and ''thermal'' freeze-out,
especially for~$\Lambda$'s.

\section{Conclusions}
In this paper we have compared the results of transport (GiBUU) and hydrodynamic
calculations for the expansion of a baryon--rich hadronic fireball. The initial events
for transport calculations have been generated in accordance with locally equilibrated phase-space
distributions of hadrons. We have found significant differences in
space--time evolution of partial densities and collective velocities of hadrons
predicted by these models. {Also,} the two models
predict very different time evolution of hadron abundances.

{The present study demonstrates} that a simple picture of sequential
chemical and kinetic freeze-outs is strongly
distorted by {the decay} and regeneration
of hadronic resonances. According to our transport calculations,
an early saturation of {the} particle multiplicities in {an}
expanding fireball
occurs only for {the} total yields, which {include}
hadrons hidden in heavier resonances. This is especially important for strange hadrons.
The hydrodynamic simulations with chemically equilibrated EoS{\hsp s} strongly underestimate
total multiplicities of hyperons and kaons. We believe that {strong}
deviations from chemical equilibrium predicted by transport calculations for such hadrons
are real and they should be explicitly taken into account in future hydro--cascade calculations~\footnote
{
In relation to this discussion we would like to mention several hydrodynamic models~\cite{Hir02,Ent06,She11}
which include chemical nonequilibrium effects.
}.

However, one should be very careful in selecting and combing the corresponding
models. Obviously, a {smooth} matching of {the} models is
{only possible} if there exists a stage when both
transport and hydrodynamic models predict a similar behavior of macroscopic quantities.
On the basis of our present analysis we conclude that these models predict {an}
essentially different evolution of particle densities and velocities.
Therefore, the combined model predictions will strongly depend on the choice of the transition
hypersurface.

One should bear in mind, that in this paper we have disregarded
viscosity effects in the~hydrodynamic calculations.
These effects may be responsible for {the} differences in collective velocities
predicted by the two models.
In the future we are going to check sensitivity of the results to the choice of initial conditions.
In particular, we plan to extend our analysis to the case of nonzero
initial collective velocity of the fireball. It would be {also} interesting to perform
calculations with inclusion of relativistic mean fields.
In this way one can effectively account for soft non-binary hadronic interactions disregarded in our
present calculations.

\begin{acknowledgments}
The authors thank \mbox{U.~Mosel} for stimulating discussions, \mbox{M.I.~Gorenstein} and
\mbox{E.L.~Bratkov}\-skaya for useful remarks. This work was supported the Helmholtz International Center for
FAIR, by the DFG \mbox{grant 436 RUS 113/957/0--1} (Germany), and the grant NSH--7235.2010.2 (Russia).
\end{acknowledgments}

\begin{appendix}
\setcounter{equation}{0}
\renewcommand{\theequation}{A.\arabic{equation}}
\section{Generation of initial events}
In our calculations we assume that initial particles are located
inside a cube with dimen\-sions $|x|,|y|,|z|<L$ ($\bm{r}=0$ corresponds to the fireball center).
This cube is divided into smaller cells with same sizes $\Delta l$ along each coordinate axis. Below we choose $L=10$~fm,
$\Delta l=0.2$~fm. For each cell we determine average numbers of the $i$-th hadrons,
$\ov{N}_i=n_i\hsp(\Delta l)^3$\,, where $n_i$ is their partial density at the cell center.
The latter is calculated by integrating the equilibrium DF (\ref{fbdf}) over the 3-momentum. This calculation
is performed for all mesons, baryons and antibaryons listed in Table~\ref{tab1}\hspace*{0.5pt}.

It is assumed that for each hadronic type particle multiplicities in a given cell are
distributed in accordance with the Poisson distribution. In this case the probability to find $n$
hadrons of the type~$i$ in the cell is given by the expression
\bel{pdis}
w_n^{(i)}=\frac{\ov{N}_i^{\,n}}{n!}\exp{(-\ov{N}_i)}\hsp.
\ee
From~\re{pdis} one can see, that the probability that a given cell is empty, i.e. contains no hadrons, equals
\mbox{$P_0=\prod\limits_{i}w_0^{(i)}=\exp{(-\ov{N}_{\rm tot})}$},
where $\ov{N}_{\rm tot}=\sum\limits_i \ov{N}_i$ is the total average multiplicity of
hadrons of all types in the cell.
For our choice of parameters, $w_1^{(i)}\simeq\ov{N}_i\ll 1$ even for most abundant hadrons. Probabilities
to find more than one hadron
($n\geqslant 2$) are negligible for all cells. In our calculations we consider
only events with cells containing {no} more than one hadron.

At the first step of the initialization procedure we generate randomly particle's coordinates
$x,y,z$ assuming that each coordinate is homogeneously distributed in the inter\-val~$(-L,L)$.
Then we find the cell containing the point $(x,y,z)$ and decide whether it is empty or not. To determine this,
we choose a random number $\xi$ {with the homogeneous distribution in the interval~$[0,1]$}. If the inequality \mbox{$\xi<1-P_0$} does
not hold, the cell is considered empty (it is excluded from further treatment of a given event). In the opposite case we
consider the cell as ''filled'' and generate the type of hadron contained in it~\footnote
{
We disregard {the} presence of cells with several hadrons of different types.
}.
Following Ref.~\cite{Pet08} we assume that the relative probability of $i$-th species
equals $\ov{N}_i/\ov{N}_{\rm tot}$\,. For hadrons with nonzero isospin $I$ we randomly generate its isospin
projection $I_3$. This is done assuming the homogeneous distribution of $I_3$ in the interval $|I_3|\leqslant I$.

At the second step we generate a hadronic 3-momentum $\bm{p}$. It is postulated that the momentum distribution
of the $i$-th hadron in a given cell is proportional to the equilibrium~DF introduced in~\re{fbdf}. Normalizing this
distribution to unity, one gets the relation
\bel{mdfi}
\frac{d^{\hsp 3}w_i}{d^{\hsp 3}p}=\frac{f^{\hsp (eq)}_i}{n_i}\equiv\varphi_i(\bm{p}).
\ee
To generate particle's momentum in accordance with distribution (\ref{mdfi}), we use the well-known rejection method~\cite{Bir76}
extending it to the case of three dimensions. First, we randomly choose three components of $\bm{p}$, assuming
that each component is homogeneously distributed in the interval $[-p_{\hsp m},p_{\hsp m}]$ where $p_{\hsp m}=3$\,GeV/$c$\hsp .
Generating a new random number $\xi\in [0,1]$\,, we accept the momentum $\bm{p}$ if the inequality
\mbox{$\xi<\varphi_i(\bm{p})/\varphi_i(0)$} holds~\footnote
{
Our initialization procedure can be generalized for initial states with nonzero collective velocities. Then one should replace $n_i$
by $\gamma\hsp n_i$ and generate (instead of $\bm{p}$) the corresponding momentum~$\widetilde{\bm{p}}$ in the local rest frame.
}.

This procedure is continued for other cells until the total net baryon number of generated particles does not exceed the value
$B_{\hsp\rm tot}$ determined by the initial density profile $n(r,0)$ (see~\re{icon}). The resulting set of coordinates, momenta and types of
particles is considered as a single event. It is used as an initial condition for subsequent GiBUU simulations.
\end{appendix}

\end{document}